\def\ref{\par \noindent \hangindent=3pc \hangafter=1}
\def\spose#1{\hbox to 0pt{#1\hss}}
\def\approxlt{\mathrel{\spose{\lower 3pt\hbox{$\sim$}}
        \raise 2.0pt\hbox{$<$}}}
\def\approxgt{\mathrel{\spose{\lower 3pt\hbox{$\sim$}}
        \raise 2.0pt\hbox{$>$}}}

\font\small=cmr10 scaled\magstep0

\magnification=\magstep1
\baselineskip=0.8truecm

\def\col{\hbox{${\rm cm}^{-2}\,$}}
\def\ergps{\hbox{${\rm erg}\,{\rm s}^{-1}$}}
\def\ergpsphz{\hbox{${\rm erg}\, {\rm s}^{-1}\, {\rm Hz}^{-1}$}}
\def\ergpspmpc{\hbox{${\rm erg}\, {\rm s}^{-1}\, {\rm Mpc}^{-3}$}}
\def\ergpcmsqps{\hbox{${\rm erg}\, {\rm cm}^{-2}\, {\rm s}^{-1}$}}
\def\kmpspmpc{\hbox{${\rm km}\, {\rm s}^{-1}\, {\rm Mpc}^{-1}$}}

\centerline{HARD X--RAY EMISSION FROM EXTRAGALACTIC $IRAS$ $12\mu$m SOURCES:}

\centerline{CONSTRAINTS ON THE UNIFIED AGN MODEL}

\centerline{AND THE SYNTHESIS OF THE X-RAY BACKGROUND}

\vskip 2truecm

\centerline{\small XAVIER BARCONS}

\smallskip

\centerline{\small Instituto de F\'\i sica de Cantabria (Consejo Superior de
Investigaciones Cient\'\i ficas -}
\centerline{\small  Universidad de Cantabria), 39005 Santander, Spain}

\smallskip

\centerline{\small Electronic mail: barcons@astro.unican.es}

\medskip

\centerline{\small ALBERTO FRANCESCHINI, GIANFRANCO DE ZOTTI}

\smallskip

\centerline{\small Osservatorio Astronomico di Padova. Vicolo dell'
Osservatorio 5, 35122 Padova, Italy}

\smallskip

\centerline{\small Electronic mail: franceschini, dezotti@astrpd.pd.astro.it}

\medskip

\centerline{\small LUIGI DANESE}

\smallskip

\centerline{\small SISSA - International School for Advanced
Studies, Strada Costiera 11, 34014 Trieste, Italy}

\smallskip

\centerline{\small Electronic mail: danese@astrpd.pd.astro.it}

\medskip

\centerline{\small AND}

\smallskip

\centerline{\small TAKAMITSU MIYAJI\footnote*{\small Present address: 
Max--Planck--Institut f\"ur Extraterrestrische Physik, Postfacht
1603, D-85740 Garching bei M\"unchen, Germany}}

\smallskip

\centerline{\small Code 666, NASA/Goddard Space Flight Center, Greenbelt, 
MD 20771, USA, and}

\centerline{\small Department of Astronomy, University of Maryland, College
Park, Maryland 20742, USA}

\smallskip

\centerline{\small Electronic mail: miyaji@rosat.mpe-garching.mpg.de}

\medskip 

\vfill\eject 

\centerline{ABSTRACT} 

\medskip 

\noindent We analyze the $2-10$~keV X-ray emission of complete samples
of AGN and galaxies selected at 12$\mu$m and recently compiled by Rush,
Malkan \& Spinoglio (1993).  The content in active galaxies of the 12$\mu$m 
sample is indeed much larger and less biased against low
luminosity and `hidden' active nuclei than for samples selected at other
wavelengths.  As a necessary complement we also study the IR emissivity
of the largest sample of hard X--ray selected AGN (Grossan 1992).  Our
purposes are to probe the unified scheme of active nuclei and to
evaluate the local X--ray volume emissivity of low luminosity and
`hidden' AGN. 

Two methods are used in our analysis to search for X-ray emission in
the HEAO-1 A2 all-sky maps from the extragalactic sources contained in
the extended 12$\mu$m sample: an X-ray Intensity Distribution Analysis
(IDA) around the known positions of the sources and a
cross-correlation function (CCF) analysis between galaxy densities and
X-ray intensities in the high galactic latitude sky ($\mid
b\mid>25^{\circ}$), where the catalogue is $>98$ per cent complete.

The joint analysis of the X-ray selected and $12\mu$m selected samples
enable us to define the relationship between the X--ray and the mid--IR
emissions of Seyfert nuclei.  In particular it turns out that congruence
is found between the hard X--ray and mid--IR Seyfert 1 nuclear
luminosity functions if a non--linear relation $L_{12\mu} \propto L_{5\,
keV}^{0.8}$ is assumed with appropriate dispersion.  Although this
result is easily explained by circum-nuclear dust re-radiation as the main
source of the mid--IR emission, nonetheless it argues against the
simplest version of the unified model for active nuclei. 

We confirm that X-ray emission is basically restricted to AGN, Seyfert
1's being much more powerful emitters than Seyfert 2's.  We find no
signal of X-ray emission for the remaining galaxies selected at
$12\mu$m, not even for those having a flat far-infrared spectrum, which
would be prime candidates to host hidden AGN.  Our result on the local
volume emissivity of the Seyfert 1's is fully consistent with those
derived from studies of the hard X--ray luminosity function.  However
the 95$\%$ upper limits to the local emissivity we derive for the
Seyfert 2 class and for the most likely `hidden' AGN candidates,
severely constrain (and in some cases exclude) models of the X--ray
background based on the unified Seyfert scheme.

We also find that less than $\sim20$\% of Seyfert 1-like AGN and less
than $\sim 50$\% of Seyfert 2-like AGN can be `hidden' in the
non-Seyfert galaxy samples assuming that the X-ray to 12$\mu$m flux ratios
measured in the 12$\mu$m selected samples can be considered typical for 
these classes of objects.  In this case, the number ratio of Seyfert 2 to 
Seyfert 1 galaxies ranges from 1 to 2.

\bigskip

\noindent{\it Subject headings:} Galaxies: general, active -- Infrared:
galaxies -- X-rays: general, galaxies, diffuse radiation

\vfill\eject

\centerline{1. INTRODUCTION}

\noindent The unified Seyfert scheme (Antonucci \& Miller 1985; see
Antonucci 1993 for a comprehensive review) predicts that active nuclei
are surrounded by torus--like opaque structures of gas and dust.
Depending on the line of sight to the nuclei, the UV and optical broad
line emitting regions may be obscured or not and the objects can be
classified as broad-- or narrow--line AGN.

Two issues of the unified scheme are relevant to our study: the nature
of the IR emission and the absorption and reprocessing of the primary
X--ray nuclear emission.  In the unified picture the most natural
explanation of the near and mid--IR emission of Seyfert galaxies is
re-radiation of the primary optical--UV radiation by thermal dust in
the torus around the active nucleus (see e.g.  Pier \& Krolik 1992;
Granato \& Danese 1994).  The near and mid--IR radiation from AGN is
expected to be mildly anisotropic in this context (Granato \& Danese
1994). However, the selection bias against obscured (type 2) and
`hidden' AGN is likely to be less severe here than in the $<10$~keV
X-ray domain where the flux transmitted through the obscuring torus
will be suppresed by a factor of several orders of magnitude (Krolik,
Madau \& Zycki 1994).  The unified picture also predicts heavy
absorption in the X--ray spectra of the narrow--line AGN with respect
to the broad--line ones (see e.g.  Krolik, Madau \& Zycki 1994).  As a
matter of fact, {\it Ginga} observations revealed high absorbing
column densities ($N_{HI}\sim 10^{23} -10^{25}\, \col$) in many
Seyfert 2 galaxies (Awaki et al 1991), whereas Seyfert 1 X--ray
spectra exhibit moderate to low or even no absorption.  Active objects
surrounded by Compton-thick tori with even higher column densities are
also expected in the unified model (Antonucci 1993).

On the other hand there are evidences that the simplest version of the
unified model is not valid (Lawrence 1991; Mulchaey, Mushotzky \&
Weaver 1992; Antonucci 1993) and that effects other than the
orientation of the line of sight must be called for to explain the
difference between type 1 and type 2 AGN.  In particular the
luminosity of the central nucleus may be a crucial parameter in a more
refined model.  In this context much can be learnt from the comparison
of the IR and hard X--ray emissions of Seyfert galaxies.

The X-ray spectral properties of the `hidden' AGN are also relevant to
the problem of the origin of the of the X--ray background (XRB).
Setti \& Woltjer (1989) first suggested that highly absorbed AGN
(possibly Seyfert 2's) may account for a major fraction of the XRB in
the 3--30 keV range.  This suggestion has been recently explored by
Comastri et al.  (1995) and Madau, Ghisellini \& Fabian (1994) in the
framework of the Seyfert's unified scheme.  However, some basic
ingredients to compute the absorbed AGN contribution the the XRB, such
as their the total number and their hard local X--ray Luminosity
Function (XLF), are still quite uncertain.  No more than a dozen
type-2 AGN have been detected by the HEAO-1 A1 all--sky survey (2--10
keV) and only a very tentative XLF has been derived (Grossan 1992).
Extremely relevant information resides in the local X--ray volume
emissivity (LXVE, i.e., the total amount of X-rays produced per unit
volume in the nearby Universe), an integrated quantity which can be
more easily constrained than a detailed XLF.  The contribution of any
class of source to the cosmic X-Ray Background (XRB) is indeed
modulated by its LXVE and cosmic evolution (see for a review Fabian \&
Barcons 1992).

With the aim of probing the unified models and of estimating the
contribution to the LXVE and to the XRB by low luminosity and `hidden'
active nuclei, we analyze with some detail in this paper the X--ray
emission from a complete sample of AGN and galaxies selected at
$12\mu$m by Rush, Malkan \& Spinoglio (1993; henceforth RMS) and the
IR emission of the largest available hard X--ray flux limited AGN
sample (Grossan 1992). The latter is suitable to explore the ratio of
the X--ray to IR luminosities of the AGN because it includes objects
in a wide interval of luminosity (about 5 orders of magnitude). As for
the RMS galaxy sample, the advantadge of the $12 \mu$m selection is in
the large fraction of AGN, which amounts to $\approx 13$\% (to be
compared with a typical $\sim 1$\% of optically selected or
$60\mu$m-selected galaxy samples) and in the fact that it is not as
biased against low luminosity and `hidden' AGN as X-ray ($<10$~keV)
selected samples: galaxy emission is in general at a minimum in the 5
to 25 $\mu$m window, where `hidden' AGN are expected to emit a
significant fraction of their luminosity.

We study the LXVE of the RMS sample objects in the $2-10$~keV band.
Although the XRB intensity is unambiguously measured in this energy
band (free from galactic contamination at high galactic latitudes,
Marshall et al 1980) and all-sky maps are available (as given by the
HEAO-1 A2 experiment), the origin of a large fraction of it
($\approxgt 75$\%) remains speculative.  Exploitation of the rich
sample of AGN in the RMS catalogue allows the determination of X-ray
emissivities from active and non-active galaxies separately and the
exploration of the emissivity of possible `hidden' AGN.  This is
crucial to constrain models of the origin of the XRB based on the
unified Seyfert scheme.

Various methods have been used to infer or measure the local X-ray
volume emissivity of different classes of sources.  The most obvious
one is to directly search for X-ray emission around the positions of a
given sample of sources.  This was first tried by De Zotti et al.
(1989) by searching in the A2 X-ray maps for emission of a list of
galaxies emitting at 12$\mu$m (actually a subsample of the RMS sample
used in this paper).  Although the large beam-aperture of the A2
collimator ($3^{\circ}\times 1.5^{\circ}$) produced serious confusion
problems, these authors were able to detect and quantify the
emissivity from the Seyfert galaxies in their sample.  But probably
the currently most popular approach is to cross correlate complete
galaxy catalogues with X-ray maps covering large enough regions
(Jahoda et al.  1991, 1992; Lahav et al.  1993; Miyaji et al.  1994;
Carrera et al.  1994).  The method consists in smearing a usually
large and complete galaxy catalogue (selected at some optical or
far-infrared wavelength) with the same beam of the all-sky X-ray
survey.  A cross correlation between the beam-smeared galaxy catalogue
and the X-ray map is expected to produce a positive signal if the
catalogued galaxies have X-ray emitters associated with them.

We fully exploit in this paper the whole X-ray database of the HEAO-1
mission, including the A1 and A2 experiments. Although various other
space observatories have surveyed at X-ray energies up to 10 keV or
slightly above at fainter fluxes, HEAO-1 is currently, and will be for
some time, the only one providing spectral sensitivity well above 10
keV (and up to $\sim$60 keV) over the entire sky. We judged that a
detailed investigation based on such data would have provided unique
information about that numerous class of hard X-ray emitters predicted
by the unified picture of AGN activity.

The paper is organised as follows.  In Section 2 the data used is
presented.  It includes the RMS galaxy catalogue and its various
subsamples, the hard X-ray flux-limited AGN sample by Grossan (1992),
and finally the hard X-ray all-sky maps from the HEAO-1 A2 survey.  In
Section 3 we discuss the methods used to investigate the hard X-ray
emissivity of various IR samples, including an Intensity Distribution
Analysis (IDA) method (which generalizes the study by De Zotti et al
1989) and the CCF method which is also briefly outlined here and
specifically tailored to our purposes.  The results of the application
of these methods are presented in Section 4.  They concern the
derivation of the average X-ray to $12\mu$m luminosity ratios and the
XLF's for the various $12\mu$m galaxy samples, and estimates of the
X-ray volume emissivities.  Particular emphasis is given to Seyfert 1
galaxies, whose X-ray emission properties (in addition to the IR ones)
are particularly well known.  Discussions involving the existence of
`hidden' classes of AGN (i.e., optically classified as non-active),
and constraints on the unified AGN model and implications on the
synthesis of the XRB are given in Section 5.  Conclusions appear in
Section 6.

A Hubble constant $H_0=50\, \kmpspmpc$ and a deceleration parameter
$q_{\circ}$=0.5 are assumed.

\bigskip

\centerline {2. THE DATA}

\medskip

\centerline{\it 2.1 The IRAS $12\mu m$ galaxy sample}

\medskip

\noindent The RMS catalogue is a list of 835 high galactic latitude
($\mid b\mid > 25^{\circ}$) extragalactic sources with $IRAS$ $12\mu$m
flux $f_{12}>0.2$~Jy, which is almost completely identified (more than
98\%).  For each source, $\alpha$, $\delta$, the redshift and the
fluxes in the 4 $IRAS$ bands (12, 25, 60, 100 $\mu m$) are given.  The
sources are classified into 5 different categories: Seyfert 1 galaxies
and QSOs, Seyfert 2 galaxies, Liners, Starburst galaxies (non-Seyferts
with high FIR luminosity) and normal galaxies.  Usual tests
($V/V_{max}$ and log N-log S) show that both the whole sample and the
various sub-samples are complete.  Only the Seyfert 1 sample shows a
marginal incompleteness, which Rush, Malkan and Spinoglio (1993)
attribute to the fact that some still unidentified Seyfert 1s could
fall in the non-Seyfert samples.

Following previous studies exploiting $IRAS$ colours of galaxies to
identify IR-selected samples of AGN (de Grijp et al 1985), we have
searched for unidentified AGN in the RMS sample through a multicolor
analysis of $IRAS$ data. We plot in Figure 1 the $f_{12}/f_{100}$ vs
$f_{25}/f_{100}$ colour-colour diagram for various galaxy classes in
the RMS sample.

We see that Seyfert galaxies occupy a well-defined elongated region on
this plot, whereas that most of the non-Seyfert galaxies cluster
around a different position in the graph.  There are, however, a
number of galaxies with FIR colours deviating from the average
galaxian colour and spread over the AGN zone.  These have a relatively
larger $12\mu$m flux compared with the average galaxy, which implies a
higher fraction of hot dust emission.  According to our view (see also
Danese et al, in preparation) these sources are prime candidates to
host a hidden AGN.  We have defined an IR sample of AGN candidates by
selecting sources with $f_{12}/f_{100}<4000\ (f_{25}/f_{100})^2$, a
criterion which maximizes the segregation between Seyfert and
non-Seyfert galaxies in the colour plane of Fig.  1.  These objects
will constitute our Hot Dust Galaxy (HDG) sample.

To summarize, we have considered in this paper 5 $12\mu$m-selected
galaxy samples: Seyfert 1, Seyfert 2, normal galaxies (liners and
starburts are also included here given their small numbers in the RMS
sample), Hot Dust Galaxies and the whole sample.  We have excluded
from our analysis $10^{\circ}$ circles around Virgo ($\alpha=12^{\rm
h}\, 28^{\rm m}$, $\delta=12^{\circ}\, 40'$) and the Large Magellanic
Cloud ($\alpha=5^{\rm h}\, 24^{\rm m}$, $\delta=-69^{\circ}\, 45'$).
Table 1 (which also includes the results from our X-ray intensity
distribution analysis) shows the number of objects contained in each
subsample.

\medskip

\centerline{\it 2.2 The X-ray data: the HEAO-1 A1 AGN sample} 

\medskip

\noindent This paper is devoted to an assessment of the X-ray properties
of mid-IR selected extragalactic sources.  If one assumes that the hot
dust responsible for the $12\mu$m emission is mostly heated by the
central engine in AGN, then a roughly proportional relation is
expected between the X-ray flux ($f_X$) and the $12\mu$m flux.  In
order to test this hypothesis, we have considered the AGN sample
(mostly Seyfert 1 galaxies) studied by Grossan (1992) and including
$IRAS$ $12\mu$m data.  The sample has been derived from the HEAO-1
Large Area Sky Survey sample in the 2-20 keV band (Wood et al., 1984).
The $12 \mu$m data are from an add-scan (SCANPI) analysis of the
$IRAS$ survey.

More than 86\% of the sources
have been identified above a flux limit of 0.95 $\mu Jy$ at 5 keV. 
Ninety six emission line AGN (85 Seyfert 1 and QSOs and 11 Seyfert 2
galaxies) make up the complete X-ray sample. As shown by Turner \& Pounds
(1989) and by Nandra \& Pounds (1994) and also implied by spectral information
on the Piccinotti et al (1992)  AGN sample, classified Seyfert 1 galaxies
are characterized by values of the column density $N_{HI}$ of absorbing
material typically lower than $10^{22}\, \col$, whereas type-2 objects have 
almost invariably larger amounts of gas.  

A plot of the monochromatic 5~keV and $12\mu$m luminosities (including
IR detections and upper limits) is shown in Figure 2.  We can see there
an approximately linear relation, which however appears to break down at
both low and high luminosities.  At $L_{12}<10^{30}\, \ergpsphz$ ($L_{12}$
denotes the luminosity at $12\mu$m) the
IR flux is probably dominated by the contribution from diffuse dust in
the host galaxy, and this flattens up the relationship between IR and
X-ray luminosities.  On the contrary, a break down at the high
luminosities may be qualitatively inferred from the fact that over 40\%
of the objects with 5~keV luminosity $L_X>10^{26}\, \ergpsphz$ and
virtually all but three radio-loud AGN at $L_X>10^{27}\, \ergpsphz$
have no $IRAS$ detection at $12\mu$m (above an average flux limit of
$f_{12} \simeq 0.1$Jy), whereas for the whole sample the IR detection
rate approaches 80\%. This is not due to a flux selection artifact, since 
excluding the 14 brightest Seyfert 1s in Grossan's sample (which have
fluxes above $2 \mu$Jy@5~keV) the remaining sources exhibit a uniform
distribution in X-ray flux of the fraction of undetections to detections
at 12$\mu$m down to the limiting X-ray flux. 

A detailed, more quantitative discussion of these effects and possible 
interpretations will be given in Sections 4 and 5 below.

\medskip

\centerline{\it 2.3 The X-ray data: the HEAO-1 A2 database}

\medskip

\noindent The database used to estimate the X-ray emissivity of the
various RMS samples consists of the HEAO-1 A2 all-sky survey map
constructed with the MED and the HED \# 3 (the A2 {\it total} band). 
Although this broad band has sensitivity from 2 to 60~keV, we formally
keep using the 2-10~keV range as our reference band.  We have also made
use of the A2 {\it hard} band data sensitive to photons in the 6-60 keV
energy range (Allen, Jahoda \& Whitlock 1994). 

Since confusion is the most severe problem in
our analysis, we choose to keep only the $3^{\circ}\times 1.5^{\circ}$
collimators, for which the conversion factor is $2.1 \times 10^{-11} \,
\ergpcmsqps$ per R15 count s$^{-1}$ for a power-law spectrum with energy 
index $\alpha_X=0.7$.
The shape of the collimator is
assumed to be triangular in both axes with FWZI of $6^{\circ} \times
3^{\circ}$ (which is $3^{\circ}\times 1.5^{\circ}$ FWHM):
$$G(\vec x)={\rm Max}\left(0,1-{\mid x\mid\over
3^{\circ}}\right){\rm Max}\left(0,1-{\mid y\mid\over
1.5^{\circ}}\right)\eqno(1)$$

The all sky map is in the form of a $720 \times 720$ matrix in ecliptic
coordinates.  Measurements are separated $0.5^{\circ}$ in ecliptic
longitude and $0.25^{\circ}$ in ecliptic latitude, therefore providing a
set of highly overlapped measurements of the X-ray sky intensity.  These
datapoints are strongly dependent, but they provide a fine enough grid
for our purposes. 

\bigskip

\centerline{ 3. THE METHODS}

\medskip

\centerline{\it 3.1 The X-ray Intensity Distribution Analysis (IDA) around IR
source positions}

\medskip

\noindent Roughly speaking this method consists in the measurement of
the X-ray intensity in the positions of the sources, and a comparison of
this intensity histogram against some reference or `blank sky'
distribution.  If there is a significant difference (that we will test
with the one sided Kolmogorov-Smirnov statistic) we would have then
detected a signal of X-ray emission from the source sample.  We will
then model this difference in terms of either a constant $f_X/f_{12}$
ratio or of a $f_X$-to-$f_{12}$ non-linear parametric relation.  This
method (already used by De Zotti et al.  1989) has the advantage over an
all-sky CCF in that it does not use data from points where there are no
sources.  This is especially relevant to our case, since the total
sample has only about 0.1 sources per beam and so 9 out of 10 X-ray data
measurements only introduce noise in the CCF. 

The IDA method, however, has some subtleties that can only be properly
met by numerical simulation techniques.  Both the significance of any
putative signal and the determination of the relation between $L_X$ and
$L_{12}$ have been done by making simulations of the X-ray sky
and then performing measurements of the X-ray intensity distribution
around the positions of the sources exactly in the same way as in the
real sky.  To generate a `blank sky', we have first measured the X-ray
intensity collected by the collimator in a large number of positions at
$\mid b\mid > 25^{\circ}$ and excluding the Virgo and LMC regions.  This
is used as the parent distribution from which we draw a `blank sky'map. 
Each one of the $720\times 720$ sky elements is replaced by a random
measurement of the above list.  It is clear that this does not reproduce
the positional structure of the real sky, neighbouring X-ray intensities
being no longer correlated.  However, this simple algorithm does not
bias the one-point background intensity distribution in any way. 

The next step is to add the sources of the corresponding subsample. 
This is done by beam-smearing (according to eq.  1) the source sample
under consideration, assigning a given relation between $f_X$ and
$f_{12}$.  In that synthetic sky map, we perform the same measurement as
in the real sky X-ray map, i.e., we build a histogram with the
intensities associated to the source positions.  These intensities are
obtained, both when measuring on the real X-ray sky and on the synthetic
sky, by a bilinear interpolation of the 4 nearest measurements. 
Although this effectively sharpens the histograms (a bilinear
interpolation acts as a smoothing procedure), the method is correct as
long as the same type of measurement is performed to the data and to the
simulations. 

The resulting histograms are compared via the one-sided
Kolmogorov-Smirnov test.  A number of simulations (typically 10) are
performed for each tested X-IR relationship.  We find that the KS
statistic from the comparison of these simulations with the real data
has usually little scatter and therefore 10 simulations have been found
to be enough.  Our best fits always provide a good enough description of
the data, the rejection probability being $<50$\%.  All quoted errors
and upper limits refer to the 95\% confidence level. 

For Seyfert 1 galaxies not only the X-ray signal is stronger, but also a
complementary information from X-ray selected samples can be used.  In
this case we have also tested more general non-linear regressions of the
X-ray on the IR flux (see details in Sect.4): $$ \log L_X = a + b\ \log
L_{12}, \eqno(2)$$ $L_X$ and $L_{12}$ being the monochromatic
luminosities in $\ergpsphz$ at 5 keV and at $12 \mu$m, respectively.  In
addition to this simple one-to-one relation, we have also tested
regressions with gaussian distributions of the residuals. 

We emphasize that this technique properly accounts for the most severe
problem that we have in this analysis, i.e.  the source confusion in the
A2 beam.  Since the measurements are performed exactly in the same way
in the real and simulated sky, there is no double counting effect for
those cases where two $12\mu$m sources fall within a beamsize, no matter
whether or not sources are clustered.  This procedure does not account
for the part of the clustering effect where one or more clustered X-ray
sources, which are not themselves among the $12\mu$m sample, are within
one beam of a $12\mu$m source.  However, given the size of the beam, the
estimated excess flux from the clustering effect is one order of
magnitude smaller than the X-ray flux from the $12\mu$m sources
themselves, for any reasonable values of the volume emissivity and
clustering properties.  Even though, this is quite difficult to
reproduce in analytical terms.  A further advantage of this method is
that we are fitting a quantity (the X-ray to $12\mu$m luminosity or flux
relation) which has a direct physical meaning and requires no further
hypotheses to be translated into, e.g., the LXVE. 

\medskip

\centerline{\it 3.2 The Cross-Correlation Function (CCF)}

\medskip

\noindent We use here the zero-lag cross correlation function between
the beam smeared version of the samples introduced in Section 2.1 and the
map of the X-ray sky presented in Section 2.3.  It is defined as

$$W={\langle N_B I_B\rangle\over\langle N_B\rangle\langle I_B\rangle}
-1\eqno(3)$$ where $N_B$ is the number of sources per beam and $I_B$ the X-ray
intensity per beam.  Averages are taken over all available beams restricted to
$\mid b\mid > 25^{\circ}$ and avoiding the Virgo and LMC regions.

The interpretation of the CCF in terms of the local volume emissivity has been
presented in several papers (Jahoda et al. 1992; Lahav et al. 1993; Miyaji et
al. 1994; Carrera et al. 1994). Defining $\eta=W\langle N_B\rangle\langle
I_B\rangle$, there are two contributions expected to this number - a Poisson
contribution ($\eta_P$, due to the coincidence of having an X-ray source
associated with a catalogued source) and a contribution from source clustering
($\eta_{cl}$, this is to account for the fact that the source number density is
enhanced around the catalogued sources).  Following the above works we can
write these contributions as $$\eta=\eta_P+\eta_{cl}={1\over
4\pi}j_0\, \omega_{col}(0)(R_P+R_{cl})\eqno(4)$$ where $j_0$ is the LXVE due to
the catalogued sources, $R_P$ and $R_{cl}$ have dimensions of a length
and reflect the contributions of the Poisson and clustering terms respectively
and $\omega_{col}(\vec x)$ is the collimator correlation function defined as
$$\omega_{col}(\vec x)=\int d^2x'G(\vec x')G(\vec x'-\vec x)\eqno(5)$$ and $G$
is the beam triangle function of eq. (1).

The poisson depth $R_P$ is just the X-ray depth of the catalogue $$R_P=\int dR
P_x(R)\eqno(6)$$ where $P_x(R)$ is the X-ray selection function.  To be
consistent with the study performed in Section 3.1, we compute this selection
function in terms of the $12\mu$m luminosity function.  
The selection function is therefore (see eq (4)
in Miyaji et al. 1994) $$P_X(R)={\int_{4\pi R^2f_{lim}}^{\infty} dL_{12}
L_X(L_{12}) \Phi_{12}(L_{12})\over\int_0^{\infty} dL_{12} L_X(L_{12})
\Phi_{12}(L_{12})}\eqno(7)$$ where $f_{lim}=0.2$~Jy is the limiting
flux of the $12\mu$m sample and $\Phi_{12}(L_{12})$ the corresponding
luminosity function. 

Assuming that the source correlation function can be parametrized as a power
law $$\xi(r)=\left( r\over r_0\right)^{-\gamma}\eqno(8)$$ (values $r_0=8\,
h_{50}^{-1}$ and $\gamma=1.65$ will be used throughout as representative 
of IR selected galaxies, although there is some
uncertainty here), the clustering depth
can be written as $$R_{cl}=\langle n\rangle H_{\gamma}A_{\gamma}r_0^{\gamma}
\int d^2x {\omega_{col}(\vec x)\over\omega_{col}(0)}\mid\vec
x\mid^{1-\gamma\over 2}\eqno(9)$$ where
$H_{\gamma}=\Gamma(1/2)\Gamma((\gamma-1)/2)/\Gamma(\gamma/2)$, $\langle
n\rangle$ is the average source density per unit volume and $$A_{\gamma}=\int
dR R^{3-\gamma}P_{12}(R)\eqno(10)$$ where the $12\mu$m selection function is
$$P_{12}(R)={\int_{4\pi R^2f_{lim} }^{\infty} dL_{12}
\Phi_{12}(L_{12})\over\int_0^{\infty} dL_{12}
\Phi_{12}(L_{12})}\eqno(11)$$ The average source density can be easily
estimated from this function by noting that $$\langle N_B\rangle
=\Omega_B\langle n\rangle\int
dR R^2 P_{12}(R)\eqno(12)$$ where $\Omega_B=\int d^2x G(\vec x)$ is the
effective beam solid angle.

The selection functions (and consequently the depths $R_P$ and
$R_{cl}$) can be easily estimated when the catalogue is large.  As we
shall see later, this is not quite the case in our catalogues,
especially for the shortest subsamples.  The errors in these
quantities will in fact propagate to large errors in the estimate of
the LXVE which is given by $$j_0={4\pi W\langle N_B\rangle\langle
I_B\rangle\over\omega_{col}(0) (R_p+R_{cl})}\eqno(13)$$

\medskip

\centerline{ 4. RESULTS}

\medskip

\centerline{\it 4.1 X-ray signals for various IR galaxy populations and
the average $f_X/f_{12}$ flux ratios}

\medskip

\noindent Table 1 shows the results of the application of the IDA method
(see Sect.  3.1) to the 5 RMS subsamples (Sect.  2.1) where we have
assumed a constant $f_X/f_{12}$ flux ratio for each subsample.  Figure 3
shows the comparison between the `blank sky' intensity histograms and
the measured ones for the 5 samples defined in Section 2.1. 

A visual inspection of Fig.  3 evidentiates that Seyfert 1 galaxies
produce a large and highly significant signal, also confirmed by the
KS test.  The $f_X/f_{12}$ value found is entirely consistent with the
one previously obtained by De Zotti et al.  (1989).  Seyfert 2s also
produce a detectable effect (at a $>2\sigma$ level), but their X-ray
to $12\mu$m flux ratio is $\sim 7$ times smaller than for Seyfert 1s.
If Seyfert 2 galaxies are just obscured Seyfert 1 galaxies, at high
enough energies (where photoelectric absorption and Compton scattering
would be negligible) both of them should have similar flux ratios.  In
the unified scheme the obscuring tori are expected to be Compton thick
and therefore X-rays will only escape in the Klein-Nishina regime,
i.e., at energies above several tens of keV, which are only marginally
accessible to the A2 detectors.  Keeping that in mind we examined the
histograms in the hard band (corresponding to energies $>10$~keV), but
they are too noisy (too broad) to test for the differences with the
total band case (see Fig 4).  Indeed, based upon the high-energy
channels of the X-ray maps, we find for Seyfert 1 galaxies
$f_X/f_{12}=(1.4_{-0.8}^{+1.0})\times 10^{-6}$ and for Seyfert 2
galaxies $f_X/f_{12}=(0.3_{-0.3}^{+0.4})\times 10^{-6}$, i.e.
consistent with the results from the total band but with larger errors
and also not too far from Seyfert 1 and 2 having similar flux ratios.

On the other hand we do not find any significant signal coming from the galaxy
(non-Seyfert) sample.  The upper limit found shows that in terms of
$f_X/f_{12}$ the average normal galaxy is more than $\sim 20$ times fainter
than a Seyfert 1 and more than $\sim 2$ times fainter than the average 
Seyfert 2 galaxy. 

However, it might still be possible that a fraction of these galaxies do
produce some X-rays, but their signal is diluted throughout the sample. 
That would be the case if, for example, some Seyfert 1s were
misidentified as galaxies.  If we assume that these objects have a X-ray
to $12\mu$m flux ratio typical of Seyfert 1s, they could be at most
$\sim 6$\% of the galaxy sample, which is still 41 objects (i.e.  a
number comparable to that of the RMS Seyfert 1 sample). 

Since it is likely that these X-ray emitters in the galaxy sample have
FIR colours similar to AGN, some signal would be expected from the Hot
Dust Galaxy sample.  Table 1 shows that there is no such signal and that
the 95\% upper limit for the X-ray to $12\mu$m flux ratio is rather
small.  If all of the type 1 active objects missed would be included in this
hot far-IR sample, we can conclude that less than 7\% (i.e., less than
11 objects) of the Hot Dust Galaxies can be Seyfert 1s.  This poses an
upper limit of $20$\% to the number of unrecognized Seyfert 1 galaxies. 

We have also carried out the analysis for the whole sample, irrespective
of classification.  A signal is also detected at a significance $>95$\%
in this case, with an average X-ray to $12\mu$m flux ratio of $0.8
\times 10^{-7}$, most of which comes indeed from the active objects. 

All these results are confirmed and strenghtned by the CCF
analysis. Such results for the different sub-samples are shown in
Table 2.  The errors on the CCF (always 95\% confidence errors) are
drawn from simulations of randomly placed sources according to the
source density for each case, since bootstrap reshuffling of the list
of $(N_B,I_B)$ pairs results in a significand underestimate of the
errors.  It can be immediately seen that normal galaxies and the Hot
Dust Galaxies do not produce any significant signal in the CCF as in
the IDA approach. A signal from the Seyfert 1 and Seyfert 2 galaxies
is detected, although the signal from the whole sample is no longer
significant. This last fact has to be understood in terms of a large
chance probability of having sources coincident with high-intensity
X-ray spots when the number of sources is large enough.

\medskip

\centerline{\it 4.2 The relationship between X-ray and mid-IR emission for
Seyfert galaxies}

\medskip

\noindent As anticipated, for Seyfert 1 galaxies we are in the position
to perform quite a detailed comparison of hard X-ray and IR properties,
by exploiting both the A2 survey information on $IRAS$-selected objects
and the $12 \mu$m $IRAS$ survey data for the A1 AGN sample (Sect.  2.2). 
For both selections we have complete samples with redshift information,
so that we can easily build up local luminosity functions (LF) and
bivariate luminosity distributions.  The IR luminosity functions for the
various RMS samples have been derived by Rush et al.  (1993, see Figure
6 below).  The simplest way to estimate a local XLF for Seyfert 1
galaxies starting from the $12 \mu$m function would be to transform it
through the average $f_X/f_{12}$ flux ratio reported in Table 1, then
assuming that this factor does not depend on the luminosity and that any
intrinsic scatter around this mean value is negligible.  A comparison of
such a local XLF with those derived by Piccinotti et al (1982) and
Grossan (1992) (see Figure 5) shows that the two roughly agree at low
luminosities, but at medium-to-high $L_X$ the predicted XLF based on the
IR selection keeps significantly below the observed one. 

A similar exercise of predicting a LF at $12 \mu$m for Seyfert 1s,
starting from the observed XLF, can be done using an average regression
of $L_{12}$ vs. $L_X$.  We have fitted an average regression to the
$L_X$-$L_{12}$ data on Seyfert 1s which appear in Figure 2.  Such an
operation brings to an IR LF which also keeps below on average, and
particularly at medium-to-low luminosities, with respect to that
estimated by Rush et al (see Figure 6). 

To overcome the above inconsistencies, we have taken into account the effects
of an intrinsic (i.e. not due to observational errors)
dispersion of the residuals in the $L_X$-$L_{12}$ ratio. An obvious 
consequence of this dispersion is to rise the predicted LF, as discussed for
example in Franceschini et al. (1994). Taking this into account brings to a 
better match of the X-ray LF's, but worsens the fit of the IR ones at the 
high luminosities, where the values predicted starting from the A1 XLF now
exceed significantly the observations.

The only way we found to bring all these observables into agreement 
was to assume a non-linear average scaling of $L_X$ on $L_{12}$, as 
parametrized in eq.(2) and with $b>1$, and to assume 
an appreciable dispersion $\sigma(log[L_X/L_{12}])$ 
of the residuals around this average regression.

There are five parameters  in this model: the normalization $a$ and slope 
$b$ in eq.(2) and the dispersion $\sigma(log[L_X/L_{12}])_{IR}$ for the
IR-selected AGN, the normalization $a^{\prime}$ and 
$\sigma(log[L_X/L_{12}])_X$ of the average regression 
line for the X-ray selected AGN. We have obviously assumed that both the 
slope $b$ and the intrinsic dispersion $\sigma(log[L_X/L_{12}])$ keep roughly 
the same values for the two selections (see Cheng et al., 1984). To derive 
best-fit values of $a$ and $b$ for the RMS Seyfert 1s, given a value for 
$\sigma(log[L_X/L_{12}])$, we have performed a set of simulations as
explained in Sect. 3.1 and compared with the A2 survey intensity 
distributions. We report in Table 3 a set of results for $a$ and $b$, 
assuming $\sigma(log[L_X/L_{12}])=0.5$, which is derived from the rms 
deviations to the fit shown in Fig. 2.

Cheng et al. (1984) have shown that in the case of a linear dependence 
($b=1$) and for local objects, there is a simple relation between the average 
luminosity ratios for the two selections and the dispersion of residuals:
$$\langle \log(L_{12}/L_X)\rangle_{IR} = \langle \log(L_{12}/L_X)\rangle_X +
{3\over 2}\, \ln(10)\ \sigma^2(\log[L_{12}/L_X]), \eqno(14)$$
where $\sigma$ is the same for both selections. This implies a link
between the parameters $a$, $a^{\prime}$ and $\sigma$.
Eq.(14) is only slightly modified in the case that $b\neq 1$. In such a 
case the two dispersions $\sigma$ do not coincide, being related by
$$\sigma(log[L_X/L_{12}])_{IR} = b\ \sigma(log[L_X/L_{12}])_{X}, \eqno(15)$$
where the left-hand quantity is the dispersion of the conditioned probability
distibution of the residuals for a given $L_{IR}$ and for IR-selected objects, 
and viceversa for the right-hand quantity.

The best-fit parameter values able to reconcile the IR and X-ray LF's
are $a=-15.25$, $b=1.3$, $\sigma(\log[L_X/L_{12}])_{IR}=0.51$,
$a^{\prime}=-14.36$.  From these values and eq.(15) we infer
$\sigma(log[L_X/L_{12}])_{X}\simeq 0.4$.  An important counter-check is
to compare the normalizations $a$ and $a^{\prime}$ with eq.(14):
$a^{\prime} = a + 0.897\simeq -14.35$, indeed quite consistent with our
previous values.  So we are confident that our description of the IR-X
relationship is accurate enough. 

Figures 5, 6 and 2 report the model predictions versus observations for the 
hard X-ray (2-10 keV) and $12\mu$m LF's, and for the $L_X-L_{12}$ plot.
Particularly relevant for us, by means of a $\chi^2$-test comparison of
predicted and observed LF's in the X-rays and IR, we estimate 95\%
confidence limits on the average slope $b=1.3^{+0.1}_{-0.05}$: this
means a very significant non linearity of the average $L_X$-$L_{12}$
regression, in the sense that at the higher IR luminosities the X-to-IR
flux ratio becomes higher and higher. 
This non-linear behaviour is not only suggested by a direct inspection of 
the observed 12$\mu$m fluxes of hard X-ray AGN (Fig.2), but also more 
strongly required to reconcile the observed LF's of Seyfert 1 galaxies in 
the IR and X-rays.

The much fainter X-ray signals provided by the IDA or CCF analyses for
Seyfert 2s galaxies do not warrant the same detailed investigation being
performed for type-1 objects.  A comparison of the rough 2-10 keV XLF
based on the small Seyfert 2 A1 sample (Sect.  2.2) with predictions
based on the IR LF and on the average observed flux ratio shows that in
this case too a sizeable scatter of the residuals plays an important
role.  Matching the observed $\langle\log(L_X/L_{12})\rangle$ for IR-
and X-ray selected objects (which differ by roughly a factor of ten in
this case) implies a value $\sigma[\log(L_X/L_{12})]>\simeq 0.55$,
slightly higher than that of type-1 AGN.  The dot-dashed line in Fig.  5
corresponds to a prediction for Seyfert 2s based on the IR LF, a
constant flux ratio (as in Table 1), and this residual distribution.  We
see it consistent with the A1 XLF, within the large statistical
uncertainties.

\medskip

\centerline{\it 4.3 Constraints on the local X-ray volume emissivities}

\medskip

\noindent Figure 5 compares the observed 2-10 keV XLF's with inferences
based on IR selected samples of galaxies and AGN.  The thick lines
correspond to the predicted contribution of Seyfert 1s and 2s, as
discussed in detail in the previous section.  For the non-Seyfert
galaxies only an upper limit can be set which is also represented as
thin continuous lines. 

Inspection of Fig 5 reveals that the XLF is by far dominated over most 
of the luminosity range by Seyfert 1 galaxies, type 2 objects contributing 
only at the lowest luminosities. This agrees with the identification rates of 
the two types in the A1 AGN sample (only 8 Seyfert 2 over 96 objects).

So, the situation is rather clear in the $10^{42}-10^{46}\, \ergps$
X-ray luminosity range, where most (if not all) the XLF is accounted for
by Seyfert (mostly type 1) galaxies emitting at $12\mu$m.  Below
$10^{42}\, \ergps$ the situation is still unclear.  The number densities
of Seyfert 1 and Seyfert 2 galaxies become similar, but the non-Seyfert
galaxies might in fact start to dominate.  If the contribution of the
non Seyferts is close to their 95\% upper limit, the XLF will continue
rising below $10^{42}\, \ergps$.  However, as emphasized many times, we
do not detect any X-ray signal from these objects, and therefore this
has to be strictly taken as an upper limit. 

With the use of these XLFs for the different classes of objects, their
contribution to the LXVE can be found.  The results are listed in the last
entry of table 1.  The LXVE is clearly dominated by the Seyfert 1 galaxies and
only some small extra contribution arises from the Seyfert 2s.  The contribution
of these separate classes is more precisely found than the total contribution
by analysing the whole sample irrespective of the classification.  This shows
the importance of isolating the classes that produce and do not produce
X-rays as we have done here. 
Note that all quoted values of LXVE in Table 1 (and Table 2 below) are
{\it equivalent} emissivities in the 2-10 keV band, i.e. they are the fluxes
in that waveband corresponding to sources providing the observed signal in
the A2 maps and having power-law spectra with energy slope $\alpha_X=0.7$. 
This means that the reported values of LXVE take into account in detail of 
the A2 response function allover the 2-60 keV band.

The CCF analysis provides an alternative method to estimate the LXVE. To
extract it from the measurements of the CCF (Table 2) we have to estimate the
depths $R_P$ and $R_{cl}$ from the catalogues themselves.  Using eq. (6) 
and assuming $L_X\propto L_{12}$ the
Poisson depth can be obtained as 

$$R_P={{\cal N}\over\sum_{i=1}^{\cal N}\left( {L_i\over 4\pi
f_{lim}}\right)^{-0.5}}\eqno(16)$$ where for simplicity $L_i$ denotes the
$12\mu$m luminosity of the $i$-th source in the catalogue (which totals
${\cal N}$ sources and the sums are performed over the whole catalogue).  We
can similarly estimate the parameters $A_{\gamma}$ and $\langle n\rangle$ as 
$$A_{\gamma}={1\over 4-\gamma}{\sum_{i=1}^{\cal N}\left( {L_i\over 4\pi
f_{lim}}\right)^{{1-\gamma\over 2}}\over\sum_{i=1}^{\cal N}\left( {L_i\over 4
\pi f_{lim}}\right)^{-{3\over 2}}}\eqno(17)$$ 
$$\langle n\rangle ={3\langle N_B\rangle\over \Omega_B}\sum_{i=1}^{\cal
N}\left( {L_i\over 4\pi f_{lim}}\right)^{-{3\over 2}} . \eqno(18)$$ 

Since our catalogue lists are rather short, these numbers are subject to 
significant
uncertainties. In order to estimate them, we bootstrapped the catalogue,
measured these quantities and computed $R_P$ and $R_{cl}$. The 95\% errors
are also shown in Table 2 where it can be seen that these errors are rather
large. Another interesting feature to realize is that the clustering depth
$R_{cl}$ is always smaller than the Poisson depth $R_P$, as opposed to the
samples studied in Lahav et al. (1993), Miyaji et al. (1994) and Carrera et
al. (1995). The reason is simply that our samples are so sparse that the
probability of having two sources in the same beam for a given source
correlation function is small even for the whole sample (see source density
entry in Table 2). 

The resulting LXVE from the CCF for the different subsamples 
is shown in last entry of Table 2.  We can see that the
uncertainties introduced by the errors in the catalogue depths translate
to sizeable variations in this number.  Comparing the last column of
Tables 1 and 2 we see that the LXVE estimates obtained by the CCF method
are consistent with those obtained via the intensity distribution
method. 

However some differences in the results of the two approaches must be
noticed.  Since the CCF method assumes constant $L_X/L_{12}$ ratio, its
prediction on the LVXE of the Seyfert 1 galaxies is in agreement with
the value derived by using the IDA method under the same assumption.  The
IDA technique is more flexible, as it allows to test more complex
relationships between IR and X--ray luminosities.  On the other hand
this method makes use of a much noisy synthetic `blank sky' and as a
consequence it soon impingies on the background noise when the analysis
concerns classes of intrinsically faint objects.  This is demonstrated
by the relatively loose upper limits to the LXVE of the HDG and normal
galaxy classes derived with the IDA method.  Thus this method is
effective in studies of populations with significant emission. 

Conversely, the CCF method provides a more integrated information, but has a
lower intrinsic noise, particularly when the sample size is large enough
(galaxies and HDG's). Moreover, contrary to the IDA, it makes full use
of the topological information present in both the X-ray sky maps and the
source spatial distribution. It is then more effective in setting tighter
limits to the emissivity of classes of weakly emitting objects. This is
particularly true when the source population under test has a spatial
distribution uncorrelated with the main sources of the X-ray background flux.
In our case, a CCF applied to non-active galaxies or HDG's is weighted
preferentially by sky regions where the X-ray background noise is lower than
average. 

Then the two methods appear to provide somewhat complementary information
on source emissivities.

\medskip
\centerline{5. DISCUSSION}

\medskip

\centerline{\it 5.1 Emission properties of AGN in the infrared and hard X-rays}

\medskip

\noindent For Seyfert 1 galaxies, for which strong X-ray signals were
found, our analysis has shown that the relation between X-ray
and mid-IR emission is markedly non-linear: powerful X-ray sources are
found to be relatively weak emitters at $12 \mu$m.  This evidence came
not only by an inspection of directly measured fluxes in Fig.  2 (a
direct but noisy and non-conclusive indication, because of the wide
scatter on the $L_X-L_{12}$ plane), but even more from a comparison of
the IR and X-ray local luminosity functions, indirectly but very
significantly requiring a non-linear scaling of the two emissions. 

We have described such an effect in terms of a simple power-law regression,
as in eq. (2). It is however quite possible that the real effect is that of
an abrupt turn-off of the IR emission at the highest values of the underlying 
nuclear power which is traced by the hard X-ray flux. In addition to this, we 
also expect some contributions to the IR flux coming from the host galaxy, 
and appearing in the lowest luminosity objects. It is easy to verify, however,
that the IR contribution of the host galaxy cannot explain by itself a
non-linear behaviour for objects spanning a very large range of luminosities.

These observations impact on our understanding of the origin of the
mid-IR emission in Seyfert 1 galaxies.  Models of Seyfert 1 nuclei based
on the assumption that the hard X--ray emission is produced by inverse
Compton scattering on IR photons by relativistic electrons would predict
a roughly linear reletionship between X--ray and IR emission (see e.g. 
Band \& Malkan 1989).

In the framework of unified schemes most of the mid--IR flux is
expected to originate from the dust in the tori around the nuclei.  In
the simplest version of the scheme (for which only the orientation of
the line of sight is relevant to the classification of the AGN) we
expect that a quasi linear relation still holds, since dust is heated
by nuclear UV radiation, which in turn is likely to be proportional to
the hard X--ray luminosity. This is based on a recent analysis by
Franceschini et al (1994) which shows that $L_X\propto L_{opt}$ (and
consequently $L_X\propto L{UV}$) as opposed to previous work by Avni
\& Tananbaum (1986) who found $L_X\propto L_{opt}^{0.8}$. In any 
case, the increase of the $L_X/L_{IR}$ ratio with increasing $L_X$ can
be explained only if the torus covering factor is a decreasing
function of the total nuclear power.

Some support to this hypothesis appears to follow from the compariosn
of the 12$\mu$m and hard X--ray LFs of Seyfert 1 and 2 galaxies.  It
is apparent from figures 5 and 6 that luminous type 2 Seyferts are
much rarer than luminous type 1 Seyferts both in X--ray (see Grossan
1992) and at $12 \mu$m.  While the result in the X--ray band can be
possibly explained by an {\it ad hoc} assumption on the hydrogen
column density, the $12 \mu$m sample corroborates the conclusion that
Seyfert 2 are intrinsically fainter than Seyfert 1 galaxies.

The difference of a factor of $\sim 7$ in the average X-ray to IR flux ratios
for type 1 and type 2 Seyferts (Table 1) has to be understood in terms of the
obscuring material in Seyfert 2s being Compton thick, in which case virtually
no X-rays escape below 10~keV.  Indeed, the thick torus will be viewed 
quasi-equatorially for obscured objects, resulting also in a relative decrease
of the Mid-IR emissivity with respect to Seyfert 1s, but still the ratio
$L_X/L_{12}$ is expected to drop substantially when going from face-on
to edge-on viewing angles in this case.  However, none of these facts can
explain why $L_{12}$ is suppressed for high luminosity AGN.

A difference of a factor of $\sim 7$ in the average X-ray to IR flux
ratios for type-1 and type-2 Seyferts (Table 1) might be explained,
if all emissions are truly nuclear, only assuming that a column density
as high as $\sim 10^{25}\, \col$ covers type-2 objects. This
implausibly high $N_{HI}$ value appears to imply that a sizeable
fraction of the mid-IR emission in Seyfert 2s is not of nuclear
origin.

In conclusion we find evidences that the unified scheme must be complemented
with the additional requirement that at increasing source power the 
covering factor of the torus significantly decreases, possibly due to the
destruction of the torus by the radiation field.
This is in keeping with the result claimed by Lawrence (1991)
on the basis of the analysis of the relative number of narrow--
and broad--line AGN in the 3CR catalogue.

\medskip
\centerline{\it 5.2 The local X-ray volume emissivities} 

\medskip

\noindent 
Our global estimates of the LXVE from sources emitting at $12\mu$m,
based on the IDA method and the assumptions of a constant $f_X/f_{12}$
flux ratio and negligible scatter of the residuals, are summarized in
Table 1. For the two X-ray emitting classes, the Seyfert 1s and 2s, we
know additional details about the bivariate $L_X/L_{12}$ luminosity
distributions. Taking them into account, the estimates of the
corresponding LXVE change to $j_X \simeq (4.0\pm 1.5)\times
10^{38}\, h_{50}\, \ergpspmpc $ (Seyfert 1s) and $j_X \simeq
(0.77^{+0.2}_{-0.5})\times 10^{38}\, h_{50}\, \ergpspmpc$ (Seyfert 2s, quoted
errors always at 2 $\sigma$). Those based on the CCF analysis are
given in Table 2.

Altogether, our estimates of the LXVE are entirely consistent with
those obtained by Miyaji et al. (1994) for the $60\mu$m 2~Jy sample
using the same X-ray data ($(4.3\pm 1.2) \times 10^{38}\, h_{50}\, 
\ergpspmpc$). This is not a completely obvious result, since selection
at $60\mu$m favours actively star forming galaxies (with little
fractional content of AGN) while selection at $12\mu$m emphasizes the
AGN content of the sample. Also, the Miyaji et al (1994) correlation
analysis is dominated by clustering while our IDA and CCD are Poisson 
dominated.

Indeed we do not find evidence for X-ray emission coming from the
non-Seyfert galaxy population (which should include the contribution
of the starburst galaxies that dominate the $60\mu$m samples). The
upper limits to the equivalent local volume emissivity for non-Seyfert
galaxies (including liners and starbursts, as classified by Rush et
al, 1993) is $3.4\times 10^{38}\, h_{50}\, \ergpspmpc$ from the IDA
and of $1.0\times 10^{38}\, h_{50}\, \ergpspmpc$ from the CCF.  For
both the confidence is higher than 95\%.

Our main conclusion is that the local volume emissivity is
dominated by Seyfert 1 galaxies (and to a lesser extent by Seyfert 2). The
fact that the RMS sample contains such a large fraction of AGN 
has allowed us for the first time to directly show that the LXVE is indeed
due to a small fraction of galaxies rather than being distributed randomly
over the whole galaxy population. 

Should a fraction of the non-Seyfert galaxies contribute somewhere to the
hard X-ray luminosity function (Fig. 5), that could happen only at quite low
X-ray luminosities.

\medskip

\centerline{\it 5.3 Hidden Active Galactic Nuclei and the fraction of covered 
to uncovered AGN}

\medskip

\noindent 
It is instructive to see whether our analysis can set any interesting
upper limit upon the number density of hidden AGN (i.e., objects
classified as `normal' galaxies but hosting a low luminosity active
nucleus) on the basis of upper limits to the X-ray emission from
non-Seyfert galaxies and even better from the Hot Dust Galaxies which
should be prime candidates to host these hidden active nuclei.

Our upper limits imply that less than 20\% of the total Seyfert 1
population are missed inside the Hot Dust Galaxy sample.  We believe
this to be a firm upper limit to the fraction of non-absorbed AGN
which have been misidentified.  However, these hidden AGN are more
likely to have properties similar to Seyfert 2 galaxies. Taking the
average X-ray to $12\mu$m flux ratio for the 12$\mu$m--selected
Seyfert 2s and the 95\% upper limit on this ratio for the Hot Dust
Galaxies we can conclude that at most 50\% of these Hot Dust Galaxies
might be misidentified Seyfert 2s.  That means that at most 50\% of
the Seyfert 2 galaxies (always 95\% upper limit) could have been
misidentified as non-active galaxies if the X-ray to 12$\mu$m flux
ratio as the one measured here holds for these objects (see again
Table 1).  We can then conclude that at most 50\% of the Seyfert 2
galaxies can be hidden. Since the space densities of type-1 and type-2
Seyferts in the RMS sample are comparable, this would imply that the
global number of `covered' AGN is less than a factor of two larger
than that of the `uncovered'. This argument, however, assumes that
hidden AGN have warm far-IR spectra and X- to 12$\mu$m flux ratios
similar to those of Seyfert 2s, and then is not conclusive about the
question of the fractional importance of such population. In particular,
if there is some sort of bias which has selected in our Seyfert 2
sample those objects which are  Compton thin, as opposed to the
hidden AGN which would be surrounded by Compton thick tori, our 
X-ray to 12$\mu$m flux ratios would not apply to these last objects
and our conclusion would be meaningless. However, there is no physical
reason to believe that such a selection effect is present.

\medskip

\centerline{\it 5.4 Contribution of different classes of sources to the X-ray
background and the unified scheme}

\medskip

\noindent 
The contribution to the XRB by a class of source with volume
emissivity $j(z)$ (which we shall parametrize as $j(z)=j_0\,
(1+z)^{3+k}$, $k=0$ corresponding to no evolution) is
$$I={cH_0^{-1}\over 4\pi}\int dz \, (1+z)^{-5} (1+\Omega_0z)^{-{1\over 2}}\,
j(z)\, K(z)\eqno(19)$$ where $K(z)$ is the K-correction (or
simply $(1+z)^{-\alpha_X}$ for a power law spectrum of energy spectral
index $\alpha_X$) and $\Omega_0$ is the density parameter. This
relationship highlights on the relevance of the volume emissivity to
computations of the background intensity contributed by different
classes of sources.

While at soft energies $E<2\, {\rm keV}$ either resolved (mainly AGN)
or resolvable sources can account for $\approxgt 70$\% of the XRB
(Hasinger et al 1993; Barcons et al 1994) and they might in fact
saturate it according to new ASCA measurements of the soft XRB
intensity (Genderau et al. 1994), these sources probably yield $<
20\%$ of the 2--10 keV XRB (Franceschini et al. 1993).  These facts
suggest the possibility that the sources dominating the hard XRB are
silent at soft energies. In this vein, first Setti \& Woltjer (1989)
presented a model for the XRB based on highly absorbed AGN such those
predicted by the unified scheme.

More detailed models based on the X--ray properties of unified schemes
of AGN have been recently discussed by Comastri et al (1995) and Madau,
Ghisellini \& Fabian (1994).  Both models assume that unobscured and
obscured AGN evolve in luminosity $L_X(z)=L_X(z=0)\times (1+z)^k$ with
$k\simeq 2.5-2.8$.  Taking into account the detailed response function
of A2 we estimate the {\it equivalent} LXVE (see sect 4.3) of obscured
objects with $N_{HI}\geq 10^{22.5}\, \col$ implied by the model of
Comastri et al (1995) amounts to about $5\times 10^{38}\, h_{50}\,
\ergpspmpc$ and to about $4.5\times 10^{38}\, h_{50}\, \ergpspmpc$ for
the model proposed by Madau et al (1994).  Note that virtually no
classified Seyfert 1 (or broad line object) has an absorbing column
larger than $10^{22.5}\, \col$ (Nandra \& Pounds 1994, Turner \& Pounds
1989) whereas almost all type 2 (or narrow line) objects display larger
$N_{HI}$ values. 

The LXVE values of obscured objects in both models exceed by more
than a factor of two the limit we computed summing up the $95\%$ upper
limits to the {\it equivalent} overall emissivity (LXVE $\simeq
2\times 10^{38}\, h_{50}\, \ergpspmpc$) we have derived for Seyfert 2,
HDGs and normal galaxies.

We believe that our limit is robust since the mid--IR selection should
not be biased against X--ray sources (AGN or galaxies), even if strongly
absorbed.  Moreover the spectral sensitivity of the HEAO-1 A2 experiment
well above 10 keV allows an exceptional capability to detect X-ray
signals from the heavily extinguished (`hidden') AGN implied by the
unification models.  As previously admitted this does not invalidate the
unification scheme since high enough absorption would reduce the
hard X-ray flux. 


However the constraint on models of the origin of the XRB is important,
as it implies that the local population of absorbed or `hidden' AGN is
on average intrinsically too faint to yield the hard XRB if mild
evolution is assumed as in the models of Comastri et al (1995) and Madau
et al (1994).  One possible alternative is to assume faster cosmological
evolution for both unobscured and obscured AGN.  But even assuming for
both classes the fast evolution rates($k=3.2-3.5$) found by Franceschini
et al (1994) to fit the soft X-ray source counts and all the related
statistics, would not solve the problem. 

A further option would be that the evolution of `hidden' AGN is faster
than that of uncovered ones.  In the unified scheme the faster
evolution may be interpreted as the consequence of the increase of the
average covering factor with increasing redshift.

Spectral evolution of the whole hard X-ray AGN population with an average 
column density increasing at increasing redshift as $\log N_{HI}\sim 22
+1.5(z/3.5)^{0.5}$ has indeed been invoked by Franceschini et al (1993)
to explain the hard XRB and the soft and hard X-ray counts.  This option
may be inserted in the framework of the unified models as a case in which
the average optical depth of the tori is decreasing with increasing cosmic
time.

\medskip

\centerline{6. CONCLUSIONS}

\medskip

\noindent The extended galaxy $12\mu$m sample has proven to be extremely
helpful in the study of the X-ray emission of extragalactic objects.  The
fact that most X-ray emitters are also $12\mu$m sources shows the importance
of the study of the RMS sample. Morever the spectral sensitivity of
of the A1 and A2 experiments of the HEAO-1 
mission have provided unique information about that numerous class of hard 
X-ray emitters predicted by the unified picture of AGN activity.

Our main conclusions can be summarized as follows:

\item{1.} Most of the local X-ray volume emissivity comes from Seyfert galaxies
and in particular from Seyfert 1 galaxies.

\item{2.} The observed X-ray luminosity function at $10^{42}\, \ergps <
L(2-10~{\rm keV})<10^{46}\, $\break $\ergps$ can be mostly accounted
for by $12\mu$m emitting AGN, under the assumption that a non--linear
relation $L_{12} \propto L_X^{0.8}$ with dispersion $\langle
\sigma[\log(L_X/L_{12})]\rangle \simeq 0.52$ holds.

\item{3.} The unified AGN scheme must be complemented with the additional
item that the covering factor decreases with increasing source power.

\item{4.} Not many hidden AGN can exist unless their $12\mu$m is suppressed by
some unexpected mechanism. There are at most 20\% of hidden Seyfert 1 galaxies
and at most 50\% of hidden Seyfert 2 galaxies.

\item{5.} Presently available models of the XRB based on the unified
scheme predict LXVEs much larger than that implied by our analysis
of the hard X--ray data. However there are viable alternatives within
the unified scheme. 

In conclusion, we suggest that both the unified scheme of AGN activity and
the proposed models for the synthesis of the XRB based on this scheme need
to be complemented, in the former case by a luminosity dependence and
in the later case by a redshift dependence.

\bigskip

\noindent We thank M. Persic and K.  Jahoda for assitance in the use of 
the HEAO-1 A2 X-ray data.  TM is grateful to Elihu Boldt for
discussions. XB, AF, GDZ and LD acknowledge support by the European
Union `Human Capital and Mobility' program, under contract
CHRX-CT92-0033.  XB is partially supported by the DGICYT under project
PB92-0501.

\vfill\eject

\centerline{ REFERENCES}

\bigskip

\ref Allen, J., Jahoda, K., Whitlock, L., 1994, Legacy, 5, 27

\ref Antonucci, R., 1993, ARA\&A, 31, 473

\ref Antonucci, R.R.J., Miller, J., 1985, ApJ, 297, 621

\ref Avni, Y., Tananbaum, H., 1986, ApJ, 305, 83

\ref Awaki, H., Koyama, K., Inoue, H., Halpern, J.P., 1991, PASJ, 43, 195

\ref Band D.L., Malkan, M.A., 1989, ApJ, 345, 122

\ref Barcons, X., Branduardi-Raymont, G., Warwick, R.S., Fabian, A.C., 
Mason, K.O., \break McHardy, I.M., Rowan-Robinson, M., 1994, MNRAS, 268, 833

\ref Boldt, E.,  1992, In: {\it The X-ray Background}, eds. Barcons, X.,
Fabian, A.C., Cambridge University Press, p. 115

\ref Boyle, B.J., Griffiths, R.E., Shanks, T., Stewart, G.C., Georgantopoulos,
I.,  1993, MNRAS, 260, 49

\ref Carrera, F.J. et al., 1993, MNRAS, 260, 376

\ref Carrera, F.J., Barcons, X., Butcher, J.A., Fabian, A.C., Lahav. O.,
Stewart, G.C., Warwick, R.S.,  1995, MNRAS, in the press

\ref Cheng, F.Z., Danese, L., De Zotti, G., Lucchin, F.,  1984, MNRAS, 208, 799

\ref Comastri, A., Setti, G., Zamorani, G., Hasinger, G., 1995, A\&A, in the 
press

\ref Danese, L., Toffolatti, L., Franceschini, A., Mart\'\i n-Mirones, J.M., De
Zotti, G., 1993, ApJ, 412, 56

\ref de Grijp, M., Miley, G., Lub, J., de Jong, T., 1985, Nat, 314, 240

\ref De Zotti, G., Danese, L., Franceschini, A., Persic, M., Toffolatti, L.,
1989, In: {\it Two Topics in X-ray Astronomy}, eds. Hunt, J., Battrick, B., ESA
SP-296, p. 737

\ref Fabian, A.C.,  Barcons, X.,  1992, ARA\&A, 30, 429

\ref Franceschini, A., Mart\'\i n-Mirones, J.M., Danese, L., De Zotti, G., 1993, 
MNRAS, 264, 35

\ref Franceschini, A., La Franca, F., Cristiani, S., Mart\'\i n-Mirones, 
J.M., 1994, MNRAS, 269, 683  

\ref Gendreau K.C., Bautz, M.W., Ricker, G.R., et al., 1995, PASJ, in the press

\ref Granato, G. \& Danese, L., 1994, MNRAS, 268, 235

\ref Grossan, B.A., 1992, Ph. D. Thesis, Massachusetts Institute of Technology

\ref Hasinger, G., Burg, R., Giacconi, R., Hartner, G., Schmidt, M., Tr\"umper,
J., Zamorani, G., 1993, A\&A, 275, 1

\ref Jahoda, K., Lahav, O., Mushotzky, R.F. \& Boldt, E., 1991, ApJ, 378, L37

\ref Jahoda, K., Lahav, O., Mushotzky, R.F. \& Boldt, E., 1992, ApJ, 399, L107
({\it Erratum})

\ref Krolik, J.H., Madau, P., Zycki, P.T., 1994, ApJ, 420, L57

\ref Lahav, O., Fabian, A.C., Barcons, X., Boldt, E., Butcher, J., Carrera,
F.J., Jahoda, K., Miyaji, T., Stewart, G.C., Warwick, R.S., 1993, Nat, 364, 693

\ref Lawrence, A., 1991,  MNRAS, 252, 586

\ref Madau, P., Ghisellini, G., Fabian, A.C., 1994, MNRAS, 270, L17

\ref Marshall, F.E., Boldt, E.A., Holt, S.S.,  et al., 1980, ApJ, 235, 4

\ref Miyaji, T., Lahav, O., Jahoda, K., Boldt, E., 1994, ApJ, 434, 424

\ref Mulchaey, J.S.,  Mushotzky, R.F., Weaver, K.A., 1992, ApJ, 390, L69

\ref Nandra, K., Pounds, K.A., 1994, MNRAS, 268, 405

\ref Piccinotti, G., Mushotzky, R.F., Boldt, E., Holt, S.S., Marshall, F.E.,
Serlemitsos, P.J., Shafer, R.A., 1982, ApJ, 253, 485

\ref Pier E.A., Krolik, J.H., 1992, ApJ, 401, 99

\ref Rush, B., Malkan, M.A.,  Spinoglio, L.,  1993, ApJS, 89, 1

\ref Setti. G., Woltjer, L., 1989, A\&A, 224, L21 

\ref Turner, T.J., Pounds, K.A., 1989, MNRAS, 240, 833

\ref Wood, K., Meckins, J., Yentis, D., et al, 1984, ApJS, 56, 507

\vfill\eject

\centerline{FIGURE CAPTIONS}

\bigskip

\noindent FIG 1.-- Far-Infrared colour-colour diagram for the RMS sources. 
Filled and open squares 
represent Seyfert 1 and 2 galaxies, respectively. Sources
included in the Hot Dust Galaxy sample are marked with a triangle. Normal 
galaxies, starburst and narrrow emission-line galaxies occupy a region
confined around $f_{100}/f_{12}\sim 30$ and $f_{100}/f_{25}\sim 20$

\medskip

\noindent FIG 2.-- The $12\mu$m versus 5~keV monochromatic luminosity relation
for the sample presented by Grossan (1992).  Filled squares show actual data
whilst triangles represent 3$\sigma$ upper limits to the $12 \mu$m flux. 
The line is a non-linear fit, estimated using {\it survival} analysis
techniques: $\log L_X=-15.25 +1.3\log L_{12}$. 

\medskip

\noindent FIG 3.-- Total band X-ray intensity histograms around the
various subsamples used (Seyfert 1, Seyfert 2, Hot Dust Galaxies,
galaxies and whole sample).  The hatched histogram represents the
control sample histogram (i.e., blank sky measurements with the same
geometry) for each case. 

\medskip

\noindent FIG 4.-- Hard band X-ray intensity histograms around the Seyfert 1
and Seyfert 2 subsamples. The hatched histogram represents the control sample
histogram (i.e., blank sky measurements with the same geometry) for each case.

\medskip

\noindent FIG 5.-- X-ray luminosity functions. Squares are from the sample by
Grossan (1992), triangles by Piccinotti et al (1982).  The hollow
circles show the contribution from Seyfert 2 galaxies estimated from
the A1 sample.  The thick continuous line is the XLF predicted from
the Seyfert 1 LF at $12 \mu$m and a bivariate luminosity
distribution, as discussed in Sect. 4.2.  The dot-dashed line is the
prediction for Seyfert 2s galaxies.  The two thin continuous lines
mark the 95\% upper limit for the contribution of the normal galaxies
and Hot Dust Galaxies.

\medskip

\noindent FIG 6.-- The $12 \mu$m local luminosity function for Seyfert 
galaxies (filled circles), estimated from the IR sample of Rush et al. 
(1993). The LF of normal galaxies is shown for comparison. The continuous 
line shows a prediction based on the 2-10 XLF of Seyfert 1 estimated from
the A1 sample (Grossan, 1992). See Sect. 4.2 for more details.

\bye